\documentclass[12pt,preprint]{aastex}

\shorttitle{Compact radio sources in M17}
\shortauthors{Rodr\'\i guez et al.}

\begin{document}

\title{Compact radio sources in M17}

\author{Luis F. Rodr{\'{\i}}guez\altaffilmark{1,2},
Ricardo F. Gonz\'alez\altaffilmark{1},
Gabriela Montes\altaffilmark{3},
Hassan M. Asiri\altaffilmark{2},
Alejandro C. Raga\altaffilmark{4}
and Jorge Cant\'o\altaffilmark{5}}


\email{l.rodriguez@crya.unam.mx}


\altaffiltext{1}{Centro de Radiostronom\'ia y Astrof\'isica, Universidad Nacional Aut\'onoma de Mexico,
Apdo. Postal 3-72, Morelia, Michoac\'an 58089, Mexico}
\altaffiltext{2}{Astronomy Department, Faculty of Science, King Abdulaziz University, 
P.O. Box 80203, Jeddah 21589, Saudi Arabia}
\altaffiltext{3}{Instituto de Astrof\'\i sica de Andaluc\'\i a (IAA), 
CSIC, Camino Bajo de Huetor 50, 18006 Granada, Spain}
\altaffiltext{4}{Instituto de Ciencias Nucleares, Universidad Nacional Aut\'onoma de M\'exico, Apdo. 
Postal 70-543, CP. 04510, D. F., M\'exico}
\altaffiltext{5}{Instituto de Astronom\'\i a, Universidad Nacional Aut\'onoma de M\'exico, Apdo.
Postal 70-264, CP. 04510, D. F., M\'exico}

\begin{abstract}
{The classic HII region M17 is one of the best studied across the electromagnetic
spectrum. We present sensitive, high angular resolution observations made
with the Jansky Very Large Array (JVLA) at 4.96 and 8.46 GHz that reveal
the presence of 38 compact radio sources, in addition to the well
known hypercompact cometary HII region M17 UC1. For this last source we find that its
spectral index of value $\sim$1 is due to a gradient in opacity across
its face. Of the 38 compact radio sources detected,
19 have stellar counterparts detected in the infrared, optical,
or X-rays. Finally, we discuss the nature of the radio emission from 
the massive binary system CEN 1a and 1b, concluding that
both are most probably non-thermal emitters, although the first is strongly time
variable and the second is steady.}
\end{abstract}


\section{Introduction}

The centimeter continuum radiation from classic HII regions is dominated by strong free-free emission 
from the extended ionized gas present there. However, when observed with the high angular
resolution provided by an interferometer, the extended emission is filtered out and one starts
to detect compact sub-arcsecond sources of various natures (see Garay et al. 1987; Churchwell et al. 
1987; Felli et al. 1993; Zapata et al. 2004 for the case of Orion A). 
The brightest of these sources are the hypercompact (HC) HII regions, that
trace the ionized gas produced by young OB stars still embedded in dense molecular gas
(e.g. Kurtz 2005; Lizano 2008). 
The externally ionized globules are also sources of free-free radiation and result
from the interaction of the UV photons of the OB stars in the region with remaining
blobs of neutral gas existing inside the HII region
(e.g. Garay et al. 1987). The proplyds (O'Dell et al. 1993) are similar to the externally ionized globules,
but in this case the object being ionized is a protoplanetary disk surrounding a
young star. The last two known types of free-free emitters are the jets emanating
from accreting protostars (Anglada 1996;
Eisloffel et al. 2000) and the spherical ionized winds produced by massive stars
(e.g. Bieging et al. 1989; Contreras et al. 1996). There are also two 
types of sources with non-thermal radio continuum emission.
Young low-mass stars can have strong magnetospheric activity and emit detectable
gyrosynchrotron radiation (Andr\'e et al. 1988). Finally, there is also strong evidence
that systems of massive binary stars can produce synchrotron radiation in the 
region where their winds collide (e.g. Pittard et al. 2006; Ortiz-Le\'on et al. 2011). 
In Table 1 we present a summary
of the characteristics of these different types of compact radio sources.

In this paper we present a 
sensitive, high angular resolution study made with the Jansky Very Large Array (JVLA) 
of the National Radio Astronomy Observatory (NRAO)\footnote{The NRAO
is operated by Associated Universities
Inc. under cooperative agreement with the National Science Foundation.}
toward the classic HII region M17 (the Omega Nebula, NGC~6618,
W38, S45). In \S~2 we present the observations, while in \S~3 we list and
briefly discuss the compact radio sources detected.
In \S~4 we use our data to present an explanation for the spectral index
of order 1 observed in the hypercompact HII region M17 UC1. In \S~5 we discuss
the time variable sources in our sample, while in \S~6 we concentrate on
CEN 1a and 1b, the members of the massive binary system that
ionizes most of M17. In \S~7 we try to model the time-variable
emission of CEN 1a in terms of a thermal model, concluding that this
is not feasible. Finally, \S~8 presents a brief discussion on some
of the other individual sources and in \S~9 we summarize our conclusions. 

\section{Observations}

The JVLA observations were made at 4.96, 8.46 and 22.46 GHz using two intermediate frequency (IF) bandwidths
of 128 MHz each, separated by 128 MHz, and containing both circular polarizations. Each IF
was split into 64 channels of 2 MHz each. For the continuum images we averaged the central
54 channels, rejecting five channels at each end of
the bandwidth.  We observed in three epochs during the year 2011:
June 16 (2011.46) and 27 (2011.49), and July 28 (2011.57). This cadence was adopted
with the purpose of searching for variability at 4.96 and
8.46 GHz in the timescales of about 10, 30, and 40 days.
The observations at 22.46 GHz were made only in the second epoch (2011 June 27) with the
purpose of determining better spectral indices.
In addition to these timescales, we searched for rapid variability (i.e. in timescales
of minutes) during the two hours of each individual run.
At all 3 epochs, the JVLA was in the highest angular resolution A configuration.

The data reduction was made using the software package AIPS of NRAO, following the 
recommendations for JVLA data given in Appendix E of its CookBook 
(that can be found in http://www.aips.nrao.edu/cook.html). The amplitude calibrator was
J1331+3030, with adopted flux densities of 7.392, 5.205 and 2.519 Jy at 4.96, 8.46 and 22.46 GHz,
respectively. The phase calibrator was J1832-1035, with the bootstrapped
flux densities given in Table 2.

The data were self-calibrated in amplitude and
phase and we made images using only visibilities with baselines larger than 50 k$\lambda$, suppressing
structures larger than $\sim 4''$. The search for variability within each individual run
was made without self-calibration, since this procedure tends to smooth out amplitude variations
within the time interval where it is applied.
At 4.96 and 8.46 GHz we analyzed regions of $8' \times 8'$ in solid angle, while at
22.46 GHz the region analyzed was $2' \times 2'$. At 22.46 GHz our 4-$\sigma$ sensitivity
at the center of the field was $\sim$0.4 mJy beam$^{-1}$ and the only source detected
was M17 UC1, a well-known hypercompact HII region embedded in the
molecular cloud adjacent to the SW of the M17 HII region
(Felli et al. 1980). At 4.96 and/or 8.46 GHz we
detected a total of 38 compact sources, in addition to M17 UC1. These
sources are listed in Table 3. The rms noise of the images
is not constant across all the solid angle analyzed for two reasons. First, the
primary beam response decreases as we move away from the center of the field (e.g. Cotton
\& Perley 2010). The correction for this effect increases both signal and noise.
In addition, in this region there is an arc-shaped ionization front to the east and northeast 
of M17 UC1 that makes it noisier than expected (Felli et al. 1984; Johnson et al.
1998). This structure most probably traces
the interaction of the ionizing photons of NGC 6618, the cluster ionizing M17,
with the molecular clump where M17 UC1 is embedded.

Assuming a typical noise of 0.05 mJy beam$^{-1}$ at the center of the field
for the 8.46 GHz observations and following Anglada et al. (1998), we expect to detect only 
about one background source above 4-$\sigma$ over all the solid angle analyzed.
We can then conclude that practically all the detected sources are related to M17. 

\section{Compact radio sources}

In Table 3 we list the 38 detected compact radio sources, giving their positions, flux densities at
8.46 and 4.96 GHz, the resulting spectral index, and their angular size.
Most of the sources are unresolved ($\leq 0\rlap.{''}2$).
In Table 4 we list the 19 compact radio sources with reported counterparts at
other wavelengths (in all cases, stars detected previously at
optical, near-infrared, or X-rays). A counterpart was considered to be associated
with the JVLA source if its position coincided within $\leq 1''$
of the radio position. We tried to find systematic differences between the radio
sources with and without counterparts by plotting them in a spectral index versus flux density diagram
(see Fig. 1). This diagram does not segregate the two classes of source and we tentatively
suggest that the radio sources without counterparts are similar to those that
have a counterpart, but are very heavily obscured which makes detection difficult at
wavelengths other than radio.

\section{The origin of the spectral index of the hypercompact HII region M17 UC1}

This bright source has an angular size of $\sim 0\rlap.{''}4$, that at a distance
of 1.98 kpc (Xu et al. 2011) implies a diameter of $\sim$0.004 pc, falling in the
category of the hypercompact HII regions (Kurtz 2005).
Its morphology is cometary.
For this source we obtain total flux densities of 44, 107, and 194 mJy at
4.96, 8.46, and 22.46 GHz, respectively. These flux densities indicate a spectral index
of $\beta = 0.8\pm0.2$ ($S_\nu \propto \nu^\beta$), consistent with the value of $\beta \simeq 1$ 
typically found in hypercompact HII regions (Hofner et al. 1996;
Kurtz 2002; Sewilo et al. 2004). The reason for this intermediate
spectral index remains unknown. Homogeneous HII regions are expected to
show $\beta \simeq 2$ when optically thick (low frequencies) and $\beta \simeq -0.1$
when optically thin (high frequencies). In these sources the transition from optically-thick to
optically-thin ocurrs in a small frequency range. The indices of $\beta \simeq 1$
observed in hypercompact HII regions
cannot be explained as the result of observing this 
transition region since they extend over
at least an order of magnitude in frequency.
An alternative explanation is given by the hierarchical clumping model of
Ignace \& Churchwell (2004), where it is proposed that 
the radio continuum energy spectrum comes from an ensemble of spherical clumps.
In this model, for example, the spectral index of $\sim 1$ should be detected
all across the face of the nebula.

Taking advantage of the excellent quality of our 8.46 GHz data, we derived the
spectral index for M17 UC1 as a function of position. Our 4.96 GHz data does not
have enough angular resolution to clearly resolve the source and our 22.46 GHz
data is centered on the NGC 6618 cluster and M17 UC1 is located too far from the
phase center and the integration is relatively short (a total of 20 minutes on-source).
We then used for the comparison with our 8.46 GHz JVLA data, archive VLA data taken at 22.49 GHz
in 1982 March 20 under project FELL. For these archive VLA observations the 
amplitude calibrator was 1331+305, with an adopted flux density of
2.52 Jy and the phase calibrator was 1730$-$130, with a bootstrapped flux
density of 4.86$\pm$0.01 Jy. To obtain a reliable astrometry, the position
of 1730$-$130 was updated to the most recently improved value provided by the
JVLA Calibrator Manual. Both the 8.46 and 22.46 GHz observations
were made in the A configuration. To obtain comparable (u,v) coverage at both
frequencies we did not include the two inner antennas at each arm
for the 8.46 GHz observations and the three outer antennas
at each arm for the 22.46 GHz observations. The resulting (u,v) coverages are
shown in Figure 2. The images at both frequencies were restored using the
synthesized beam of the 8.46 GHz observations.

In Figure 3 we show the contour images at 8.46 and 22.49 GHz as well as
the derived spectral index from these two images. The spectral index
image shows that there is a smooth gradient from values of $\sim 2$ (optically-thick
emission) near the
head of the cometary nebula to values of $\sim -0.1$ (optically-thin emission)
near the tail. This result suggests that, at least in this hypercompact
HII region the intermediate value of $\sim$1 for the spectral index
comes from a gradient in optical depth across the face of the nebula.
We speculate that if this is the explanation for the spectral index
found in other hypercompact HII regions, they should show a
cometary morphology when observed with sufficiently high angular resolution.
For this effect to be observable at frequencies similar to those
used here, we need objects with large average emission measures,
of order $\sim10^9$~cm$^{-6}$~pc, so that they can present significant
optical depths at frequencies of order 15 GHz.

\section{Time-variable radio sources}

Of the 38 compact sources detected, two of them show large variations between the
three observing runs. The first time variable source is CEN 1a, whose flux densities are listed
in Table 5. Images at the three epochs are shown in Fig. 4. This source was not detected
in the first two epochs but in the last epoch it
became quite bright, with a flux density about an order
of magnitude larger than the 4-$\sigma$ upper limits obtained for the first two epochs.
The other variable source was the B1 star NGC 6618 B 159,
whose flux densities are given in Table 6.

In addition to possible long term variability between runs, we also searched for
short timescale variability within the individual runs. Our procedure to search for such
variability was the following. First, images of the source of interest were made with the 
AIPS task IMAGR 
for each observation epoch. With the AIPS task JMFIT, the precise position of the primary component 
was determined. The $(u,v)$ data were recentered at the position of the main component using 
the AIPS task UVFIX. With these new $(u,v)$ data, the real and imaginary parts of the interferometer 
data were 
plotted as a function of time, averaged over the $(u,v)$ plane. The real part gives us
information on the flux density of the source and the imaginary part on its position and
symmetry. We then averaged over time in bins of approximately 
10 minutes. A detailed description of this technique is given by Neria et al.\ (2010).
The flux density as a function of time for the two time variable sources
during a given run is shown in Figures 5 (CEN 1a) and 6 (NGC 6618 B 159). No fast variability
(hours or less) was detected for these sources.

Finally, of the seven sources detected by Rodr\'\i guez et al. (2009), only one
is not detected by us. This source is named VLA 2 by Rodr\'\i guez et al. (2009)
and is located at the position
$\alpha(2000) = 18^h~ 20^m~ 29\rlap.^s84;~ \delta(2000) = 16^\circ~ 10'~
15\rlap.{''}2$. Rodr\'\i guez et al. (2009) reported an 8.46 GHz flux density of
1.3$\pm$0.1 mJy, while from the images made from the concatenated (u,v) data we set a
4-$\sigma$ upper limit of 0.07 mJy. As noted before, also CEN 1a is time-variable.
The remaining five sources detected by Rodr\'\i guez et al. (2009)
were detected by us at similar flux densities.

\section{CEN 1a and CEN 1b}

CEN 1a and CEN 1b are a massive binary system (with angular separation of
$\sim 1\rlap.{''}8$; Chini et al. 1980). The O4 V stars that form this binary system
are the main ionizing sources of the 
M17 H~II region (e. g. Hoffmeister et al. 2008) and the brightest X-ray objects in the zone
(Broos et al. 2007). CEN 1a and CEN 1b were first detected as radio sources
by Rodr\'\i guez et al. (2009). Our data provides new information on these radio
sources. While CEN 1a is strongly time-variable (see Table 4), CEN 1b appears
to be a steady source. The flux densities found by us for 2011 June-July 
are consistent with the flux densities reported by Rodr\'\i guez et al. (2009)
for 1988 October and 2000 December. The available information suggests that
CEN 1b is constant over timescales of decades. Its spectral index, as derived from the
2011 June-July observations is $-0.7 \pm 0.2$, suggesting optically-thin synchrotron
emission. 

In constrast, CEN~1a  is strongly variable. As noted above, it was detected only in the
last epoch of our observations, showing a spectral index of $-0.1 \pm
0.1$. As was pointed out by Rodr\'iguez et al. (2009), the flux density
expected from a stellar wind in a single star scenario is too low  ($\sim$
0.1~mJy at 8.46~GHz) to account for the relatively large flux
density detected ($\sim$2.6 mJy at 8.46 GHz). 
This fact and the strong variability observed suggest a binary origin for
the source, with the emission arising from a wind-wind
colliding region (WCR) between the stars forming the binary system.
In this binary scenario, the variability is thought to be the result of an
eccentric orbit and/or a high inclination angle of the orbital plane.
In this context, the spectral index observed could be compatible with both
optically-thin thermal emission (Pittard 2010) and non-thermal
(synchrotron) emission (Contreras et al. 1997), arising both from
the WCR.
Its flat value is in agreement with synchrotron emission being affected by
free-free absorption from the unshocked winds, turning the negative value
expected from intrinsic optically-thin synchrotron emission into the flat
spectrum observed for CEN~1a.
The thermal contribution from a WCR is expected to increase the flux
density only a factor of a few with respect to that from the single stellar wind (Stevens
1995, Pittard 2010). On the other hand, the non-thermal emission has been
observed to increase the flux by more than one order of magnitude with
respect to the minimum states of the radio light curves of colliding wind
binaries (CWB; e.g. HD~168112; Blomme et al. 2005). Thus, the strong
increase in the flux observed in CEN~1a suggests a non-thermal origin for
the emission detected.

However, given that the spectral index is consistent with optically-thin free-free emission,
we have considered thermal emission models, but cannot find any that are reasonable and 
match the observations.
We conclude that the observed radio emission from
CEN 1a most probably has a non-termal nature and that the
observed flat spectrum is consistente with a combination of
synchrotron emission and free-free absorption, as noted before.
Non-thermal emission in massive stars is usually attributed to
them being binary systems, with the emission coming from the wind collision
region. Alternatively, the radio emission could be coming
from Fermi acceleration in wind shocks of a single star
due to the radiation-driven instability (Lucy \& White 1980;
White \& Chen 1994).
This system provides the opportunity to study simultaneously
two stars, a variable one (CEN 1a) and a steady one (CEN 1b)
and to try to understand what is the reason for their  
different behaviours.

\section{Notes on individual sources}

\subsection{JVLA 3}

This radio source is within a few arcsec of the bright infrared object KW (Kleinmann \& Wright 1973;
Povich et al. 2009) and it is probably associated with it.

\subsection{JVLA 35}

This source is remarkable because it is clearly detected at 8.46 GHz, but not
at 4.96 GHz (see Figure 7), resulting in a spectral index of
$\geq 2.9 \pm 0.6$. This spectral index is consistent with optically-thick free-free
emission, optically-thick synchrotron or even with optically thin dust emission.
The source is arc-shaped (see Figure 7) and could be tracing the edge of
a heated dust globule. There are no known counterparts to this source
in the SIMBAD database. We also searched unsuccessfully for a counterpart
in the GLIMPSE catalog of candidate young stellar objects with a high 
probability of association with the M17 complex (Povich et al. 2009).
Its unusual radio spectrum deserves further study.

\section{Conclusions}

We presented sensitive, high angular resolution observations made
with the Jansky Very Large Array (JVLA) at 4.96 and 8.46 GHz
toward the HII region M17. Our main conclusions are listed in what
follows.

1. We detected 38 compact radio sources, practically all associated
with the region. Only 19 of these sources have counterparts,
in all cases stars previously detected at infrared, optical,
or X-rays. We argue that the radio sources without counterparts
are similar to those with counterparts, but that they are heavily
obscured, which makes detection difficult at wavelengths other than radio.

2. We studied the spectral index of the hypercompact HII region
M17 UC1, finding that its spectral index of value $\sim$1 is due to a gradient in 
optical depth across
its face. We speculate that this type of gradient could explain similar
spectral indices detected in other hypercompact HII regions. In this
case, all these sources should show a cometary morphology.

3. We discussed the nature of the radio emission from 
the massive binary system CEN 1a and 1b, concluding that
both are most probably non-thermal emitters, although
the first one is highly time variable
and the second one is steady. Non-thermal emission in massive stars is usually attributed to
them being binary systems, with the emission coming from the wind collision
region. This system provides the opportunity to study simultaneously
two stars, a variable one (CEN 1a) and a steady one (CEN 1b)
and to try to understand the reasons for their  
different behaviours.

\acknowledgments
J.C., R.F.G., A.C.R., and L.F.R. acknowledge the financial support of DGAPA, UNAM and CONACyT, 
M\'exico. The National Radio 
Astronomy Observatory is a facility of the National Science Foundation operated under cooperative 
agreement by Associated Universities, Inc.

\clearpage

\begin{figure}[!t]
 \centerline{\includegraphics[height=1.0\textwidth,angle=0]{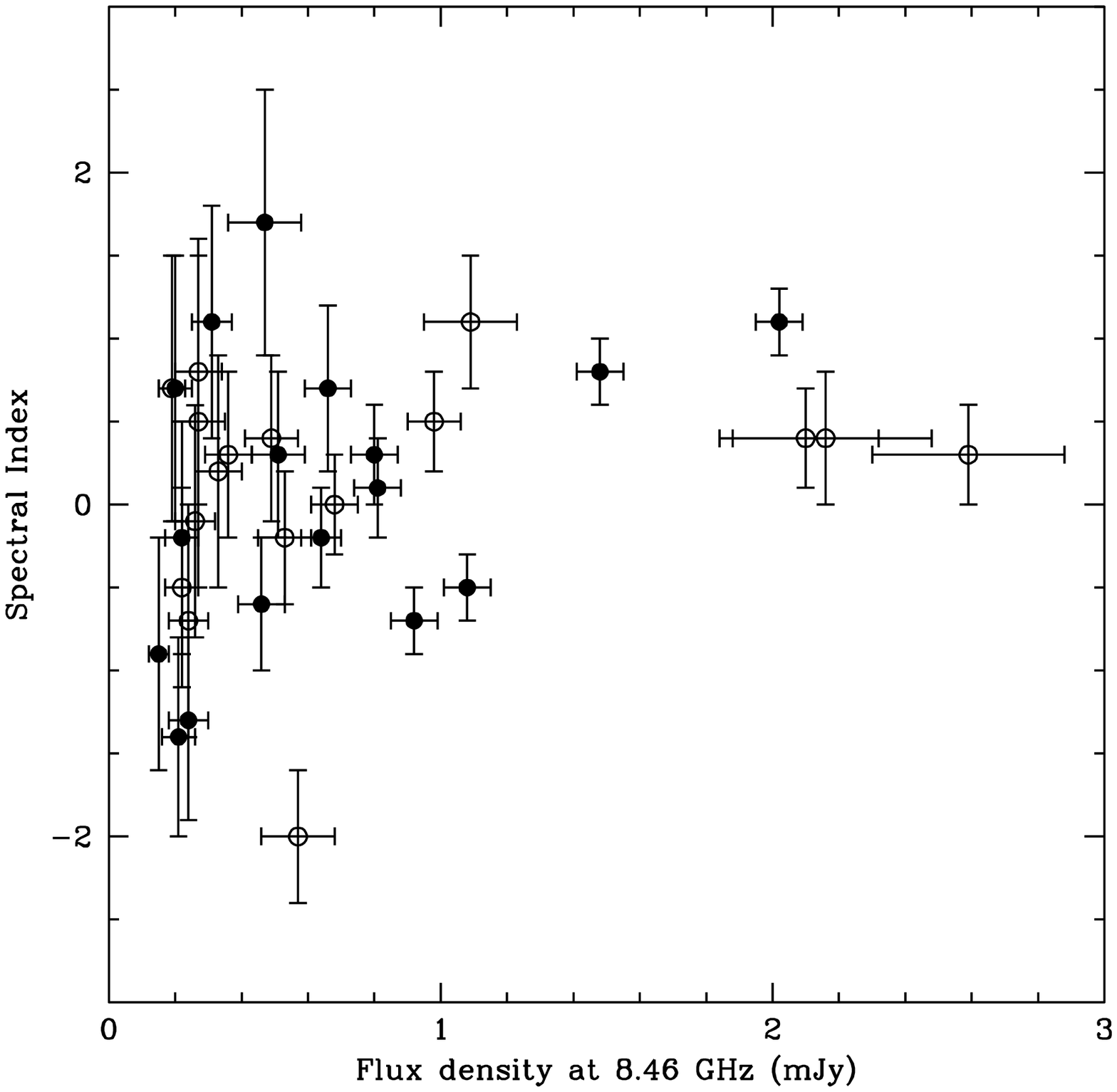}}
  \caption{Spectral index versus flux density at 8.46 GHz for the
 32 compact radio sources with determined spectral index. The
 empty circles indicate the sources without counterparts, while
 the solid circles indicate the sources with counterparts.
 \label{svsp}}
 \end{figure}

\begin{figure}[!t]
 \centerline{\includegraphics[height=0.5\textwidth,angle=0]{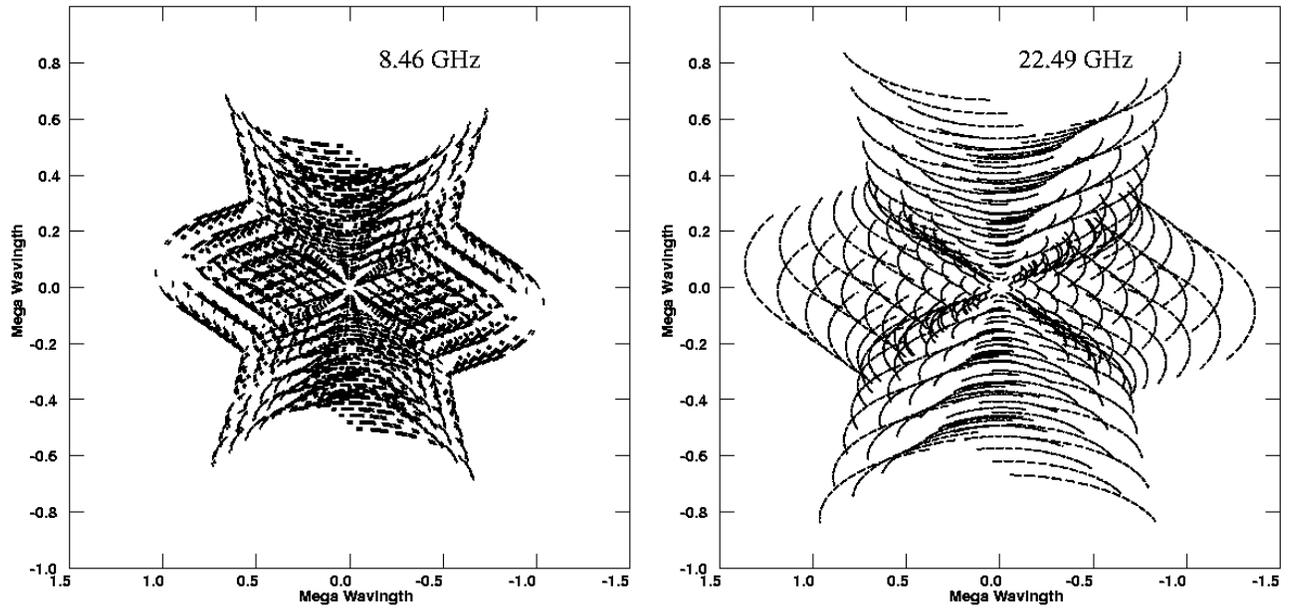}}
 \caption{(u,v) coverage of the 8.46 (left)
 and 22.49 GHz (right) observations used to estimate
 the spectral index of M17 UC1.
 \label{uvcov}}
\end{figure}

\begin{figure}[!t]
\centerline{\includegraphics[height=1.0\textwidth,angle=0]{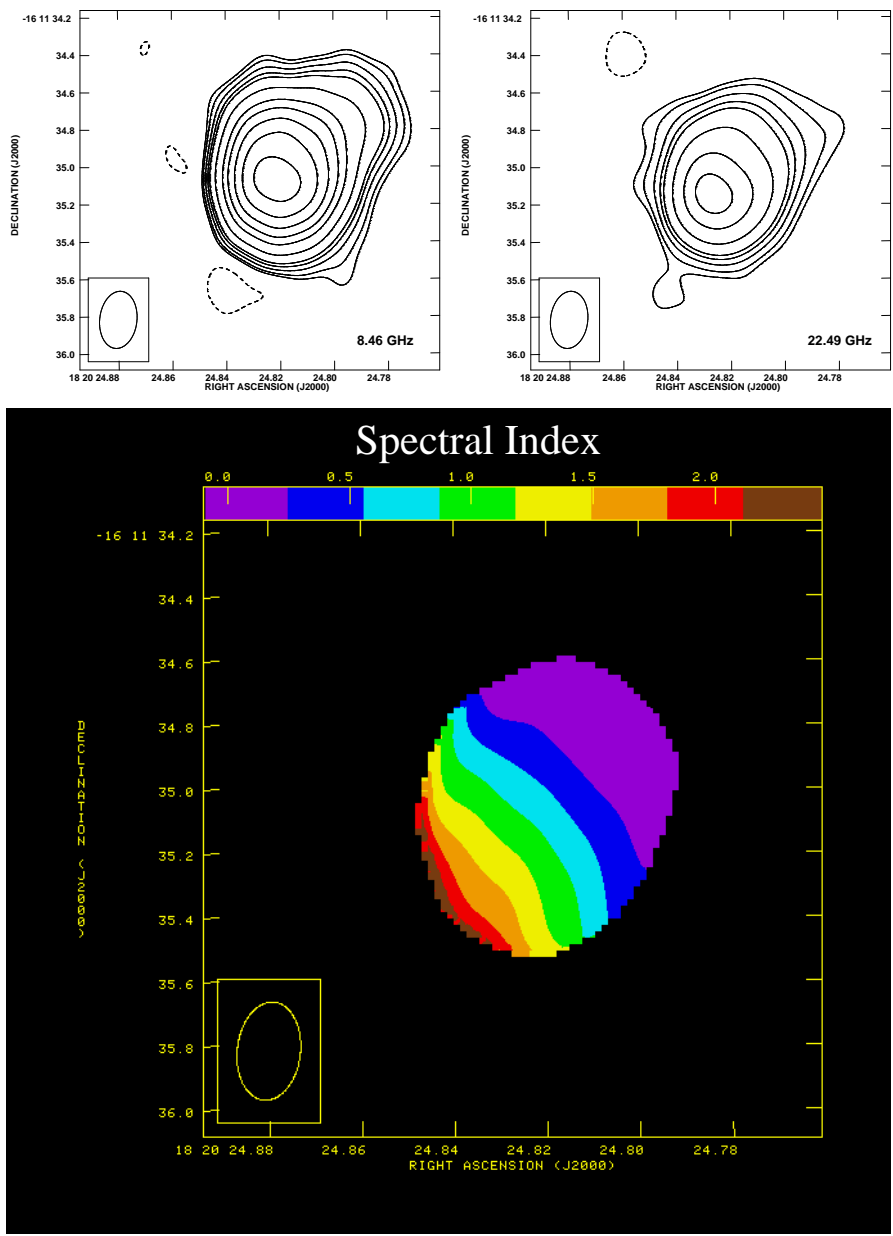}}
\caption{(Top left) JVLA contour image of M17 UC1 at 8.46 GHz. (Top right) VLA contour image
of M17 UC1 at 22.49 GHz. For these two images the contours are
-4, 4, 6, 10, 15, 20, 40, 80, 120, 180, 260, and 360 times
0.065 mJy beam$^{-1}$ (8.46 GHz) and 0.45 mJy beam$^{-1}$ (22.49 GHz),
the respective rms noise of each image. (Bottom) Color image
of the spectral index as a function of
position, as derived from the 8.46 and 22.49 GHz images. The color bar on
top of this image gives the coding for the spectral index.
The half power contour of the beam is $0\rlap.{''}31 \times 0\rlap.{''}20$ with
$PA = -7^\circ$ and it is shown in the
bottom left corner of the images.
 \label{spxk}}
\end{figure}

\begin{figure}[!t]
  \centerline{\includegraphics[height=1.0\textwidth,angle=0]{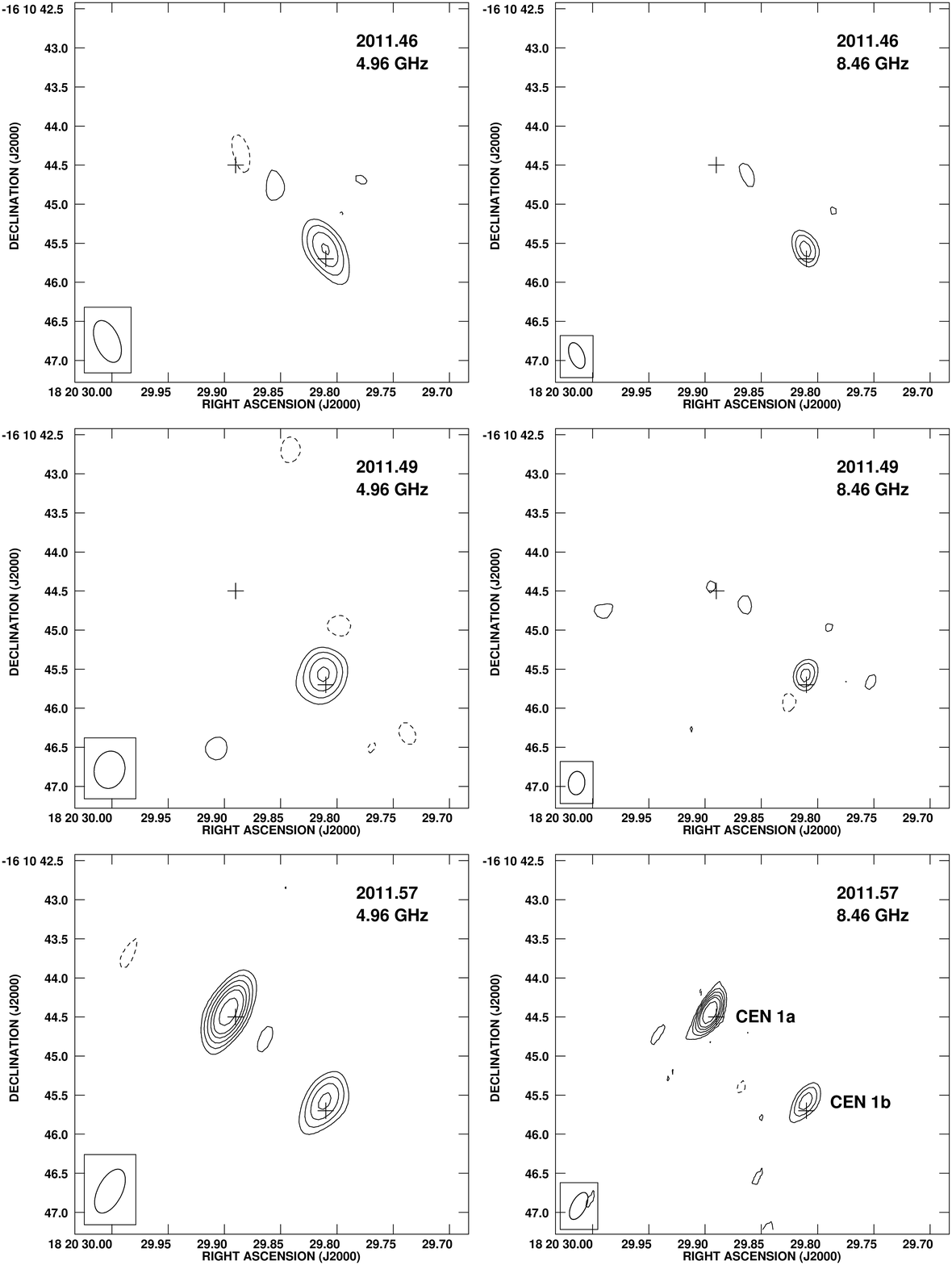}}
  \caption{JVLA images of the CEN 1 region. In each panel the crosses 
  mark the X-ray positions
 of CEN 1a and CEN 1b given by Broos et al. (2007).
(Left column) Images at 4.96 GHz for the three epochs of observation.
 Contours are $-$3, 3, 6, 10, 15, 20 and 30 times 72 $\mu$Jy beam$^{-1}$, the average rms noise
 of the images. The half power contour of the synthesized beams is shown in the 
 bottom left corner of each panel. (Right column) Same as in the left column, but for
the frequency of
8.46 GHz. Contours are given times 57 $\mu$Jy beam$^{-1}$, the average rms noise
 of the images.
\label{cen1ab}}
\end{figure}

  \begin{figure}[!t]
\centerline{\includegraphics[height=1.0\textwidth,angle=0]{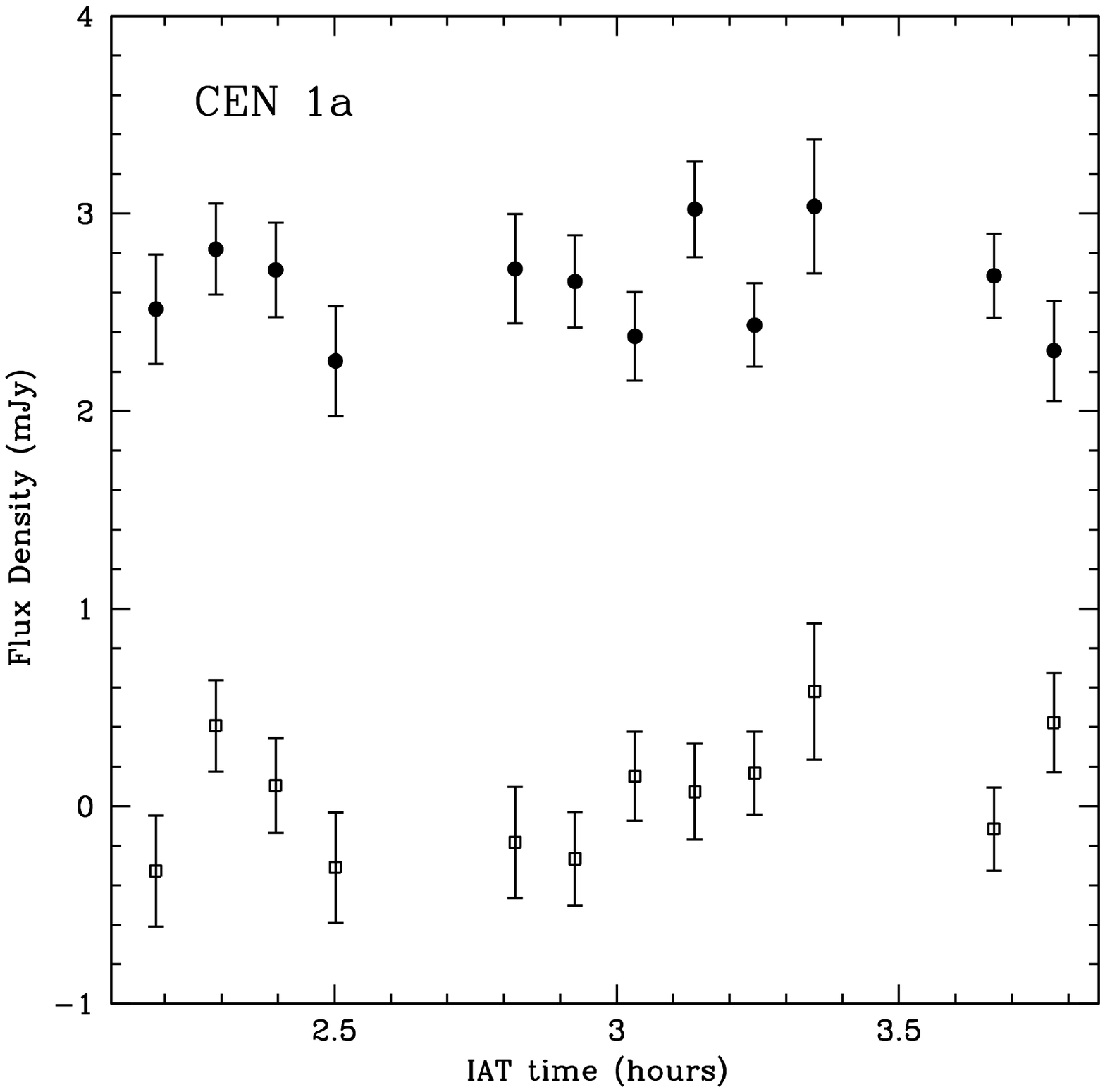}}
 \caption{Real (filled circles) and imaginary (empty squares)
 components of the interferometer data for CEN 1a at 8.46 GHz for 2011 
 July 28 as a function of international atomic time (IAT). 
 The real component is consistent with no significant variability
  over the period of the observation and a constant flux density of
  $2.6 \pm 0.1$ mJy.
 The imaginary component is consistent
with zero, indicating that the source is symmetric about
 the phase center (the origin of the visibility plane) and
 has no significant structure on these spatial scales.
 \label{fig_timex}}
 \end{figure}

\begin{figure}[!t]
 \centerline{\includegraphics[height=1.0\textwidth,angle=0]{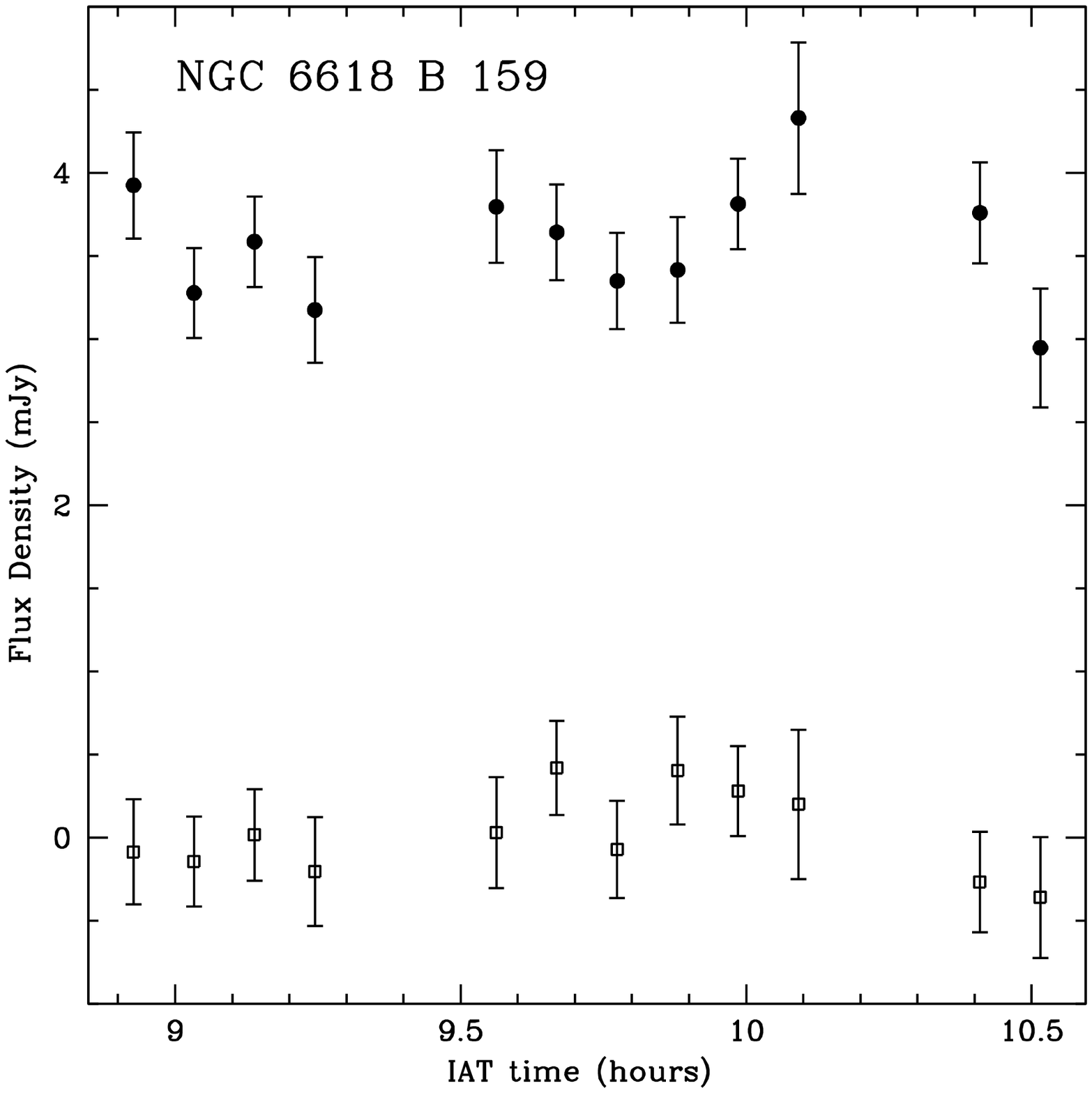}}
 \caption{Real (filled circles) and imaginary (empty squares)
 components of the interferometer data for NGC 6618 B 159 at 8.46 GHz for 2011 
 June 16 as a function of international atomic time (IAT). 
 The real component is consistent with no significant variability
 over the period of the observation and a constant flux density of
$3.8 \pm 0.1$ mJy.
 The imaginary component is consistent
  with zero, indicating that the source is symmetric about
  the phase center (the origin of the visibility plane) and
   has no significant structure on these spatial scales.
   \label{fig_timexb159}}
 \end{figure}

\begin{figure}[!t]
  \centerline{\includegraphics[height=0.5\textwidth,angle=0]{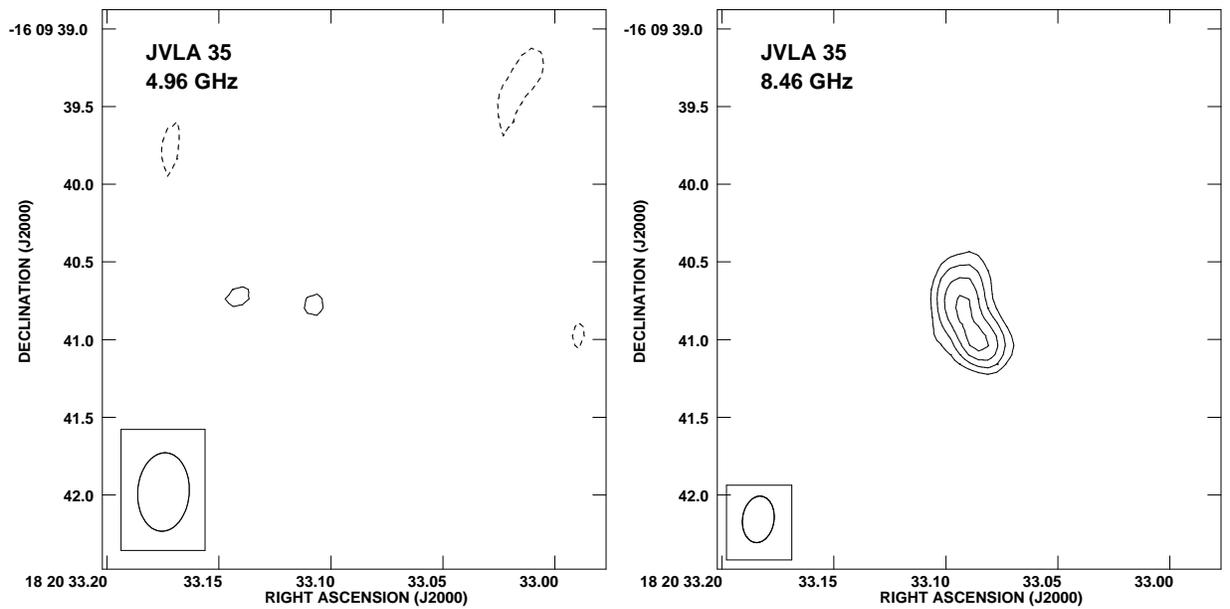}}
  \caption{JVLA images of the source JVLA 35 at 4.96 (left) and 8.46 GHz (right).
  Contours are $-3$, 3, 4, 5, and 6 times 30$\mu$Jy beam$^{-1}$.
  The source is clearly detected at 8.46 GHz, but not at 4.96 GHz.
  The half power contour of the synthesized beam is shown in the
  bottom left corner of the images and is
  $0\rlap.{''}51 \times 0\rlap.{''}33$ with
  $PA = -4^\circ$
  for the
  4.96 GHz image and 
  $0\rlap.{''}30 \times 0\rlap.{''}20$ with
  $PA = -8^\circ$ for the
  8.46 GHz image.
  \label{jvla35all}}
  \end{figure}

\clearpage

\begin{table}
\small
\begin{center}
\caption{Characteristics of compact radio continuum sources in HII regions.\label{tbl-1}}
\begin{tabular}{ccccccc}
\tableline\tableline
Class of & Emission & Spectral &  & Time & &  \\
Source & Mechanism & Index & Polarization & Variability & Morphology & Excitation \\
\tableline
HC HII region & Free-free & $\sim$1 & No & No & Various & Internal \\
Ionized globule & Free-free & $\sim$$-$0.1 & No & No & Cometary & External \\
Proplyds & Free-free & $\sim$$-$0.1 & No & No & Cometary & External \\
Jet & Free-free & $\sim$0.6 & No & Yes & Elongated & Internal \\
Spherical wind & Free-free & $\sim$0.6 & No & No & Unresolved & Internal \\
Low-mass protostars & Gyrosynchrotron & $-$2 -- +2 & Circular & Yes & Unresolved & Internal \\
Massive binary stars & Synchrotron & $\sim$$-$0.7 & Linear & Yes & Cometary & Internal \\

\tableline
\end{tabular}
\end{center}
\end{table}

\clearpage

\begin{table}
\begin{center}
\caption{Bootstrapped flux densities for the phase calibrator J1832-1035.\label{tbl-2}}
\begin{tabular}{cccc}
\tableline\tableline
Frequency & 2011 Jun 16 & 2011 Jun 27 & 2011 Jul 28 \\
(GHz) & (Jy) & (Jy) & (Jy) \\
\tableline
4.96 & 1.359$\pm$0.003 & 1.315$\pm$0.002 & 1.314$\pm$0.002  \\
8.46 & 1.433$\pm$0.004 & 1.404$\pm$0.002 & 1.404$\pm$0.003 \\
22.46 & ---  & 1.072$\pm$0.006  & ---  \\
\tableline
\end{tabular}
\end{center}
\end{table}

\clearpage

\begin{deluxetable}{cccccccc}
\tablecolumns{8}
\tablewidth{0pc}
\tablecaption{Compact radio continuum sources in M17}
\tablehead{
\colhead{} JVLA & \multicolumn{2}{c}{Position$^a$} & 
& \multicolumn{2}{c}{Flux Density(mJy)} & Spectral & Angular \\  
\cline{2-3}  \cline{5-6} 
\colhead{} Number & $\alpha$(2000) & $\delta$(2000) & & 8.46 GHz & 4.96 GHz & Index & Size$^b$ 
}
\startdata
1 & 18 20 13.335 & -16 12 50.69 & &  $\leq$0.48$\pm$0.12 &  1.14$\pm$0.10 & $\leq$$-$1.6$\pm$0.5 &  U   \\
2 & 18 20 17.575 & -16 12 21.52 & &  0.57$\pm$0.11  & 1.63$\pm$0.13 &  $-$2.0$\pm$0.4 &  $0\rlap.{''}4$   \\
3 & 18 20 19.475 & -16 13 29.87 & &  0.33$\pm$0.06 & $\leq$0.20$\pm$0.05 & $\geq$+0.9$\pm$0.6  & U   \\
4 & 18 20 21.435 & -16 12 06.81 & &  0.21$\pm$0.05 & 0.45$\pm$0.11 &  $-$1.4$\pm$0.6  & U  \\
5 & 18 20 21.637 & -16 11 17.91 & &  0.66$\pm$0.07 & 0.46$\pm$0.10 & +0.7$\pm$0.5  &   U   \\
6 & 18 20 22.369 & -16 12 04.95 & &  0.27$\pm$0.08 & 0.21$\pm$0.09 & +0.5$\pm$1.0  &   U  \\           
7 & 18 20 25.317 & -16 11 48.74 & &  1.09$\pm$0.14 & 0.60$\pm$0.11 & +1.1$\pm$0.4  & $0\rlap.{''}3$  \\        
8 & 18 20 25.496 & -16 10 53.77 & &  0.47$\pm$0.11 & 0.19$\pm$0.07 & +1.7$\pm$0.8  & U   \\
9 & 18 20 25.297 & -16 11 39.83 & &  2.16$\pm$0.32 & 1.70$\pm$0.31 & +0.4$\pm$0.4  &  $0\rlap.{''}6$  \\        
10 & 18 20 26.239 & -16 11 15.85 & &  0.68$\pm$0.07 & 0.69$\pm$0.09 &  +0.0$\pm$0.3  &   U  \\           
11 & 18 20 26.745 & -16 07 21.82 & &  2.10$\pm$0.22 & 1.66$\pm$0.15 & +0.4$\pm$0.3 & $0\rlap.{''}4$ \\
12 & 18 20 27.377 & -16 11 38.59 & &  0.53$\pm$0.08 & 0.60$\pm$0.10 &  $-$0.2$\pm$0.4 &  U  \\           
13 & 18 20 27.855 & -16 09 55.60 & &  0.15$\pm$0.03 & 0.24$\pm$0.07 &  $-$0.9$\pm$0.7 &  U  \\
14 & 18 20 28.022 & -16 11 27.00 & &  0.36$\pm$0.07 & 0.30$\pm$0.06 &  +0.3$\pm$0.5 &  U  \\ 
15 & 18 20 28.233 & -16 13 17.83 & &  0.64$\pm$0.06 & 0.72$\pm$0.09 & $-$0.2$\pm$0.3 &  U  \\
16 & 18 20 28.283 & -16 11 30.53 & &  0.22$\pm$0.05 & 0.24$\pm$0.07 & $-$0.2$\pm$0.7  & U  \\
17 & 18 20 28.384 & -16 10 14.13 & &  0.33$\pm$0.07 & 0.30$\pm$0.09 & +0.2$\pm$0.7 &  U  \\           
18 & 18 20 29.118 & -16 08 58.61 & &  2.59$\pm$0.29 & 2.26$\pm$0.19 & +0.3$\pm$0.3  & $0\rlap.{''}6$  \\
19 & 18 20 29.436 & -16 10 49.84 & &  0.51$\pm$0.08 & 0.43$\pm$0.10 & +0.3$\pm$0.5 &  U  \\
20 & 18 20 29.811 & -16 10 45.58 & &  0.92$\pm$0.07 & 1.36$\pm$0.09 & $-$0.7$\pm$0.2  &  U  \\
21$^c$ & 18 20 29.897 & -16 10 44.45 & &  1.48$\pm$0.07 & 0.96$\pm$0.08 &  +0.8$\pm$0.2  &  U  \\
22 & 18 20 30.003 & -16 10 35.26 & &  0.81$\pm$0.07 & 0.77$\pm$0.09 & +0.1$\pm$0.3 & U  \\
23 & 18 20 30.131 & -16 10 39.45 & &  0.46$\pm$0.07 & 0.63$\pm$0.11 & $-$0.6$\pm$0.4  &  U  \\
24 & 18 20 30.323 & -16 10 50.65 & &  0.24$\pm$0.06 & 0.35$\pm$0.09 & $-$0.7$\pm$0.7  &  U  \\           
25 & 18 20 30.342 & -16 11 43.67 & &  0.26$\pm$0.06 & $\leq$0.16$\pm$0.04 &  $\geq$+0.9$\pm$0.6 &  U  \\
26 & 18 20 30.415 & -16 11 03.63 & &  0.27$\pm$0.07 & 0.18$\pm$0.06 &  +0.8$\pm$0.8  &  U  \\           
27 & 18 20 30.442 & -16 10 53.08 & &  0.24$\pm$0.06 & 0.48$\pm$0.09 & $-$1.3$\pm$0.6 &  U  \\
28 & 18 20 30.577 & -16 11 04.13 & &  0.80$\pm$0.07 & 0.69$\pm$0.08 & +0.3$\pm$0.3  &  U  \\
29 & 18 20 30.632 & -16 10 28.51 & &  0.31$\pm$0.06 & 0.17$\pm$0.05 & +1.1$\pm$0.7 &  U  \\
30 & 18 20 30.777 & -16 10 59.35 & &  0.98$\pm$0.08 & 0.74$\pm$0.08 & +0.5$\pm$0.3  &   U  \\           
31$^c$ & 18 20 31.112 & -16 09 29.83 & &  2.02$\pm$0.07 & 1.13$\pm$0.09 & +1.1$\pm$0.2 &   U  \\
32 & 18 20 31.523 & -16 10 32.56 & &  0.20$\pm$0.05 & 0.14$\pm$0.05 & +0.7$\pm$0.8 & U  \\
33 & 18 20 32.981 & -16 11 15.55 & &  0.19$\pm$0.04 & 0.13$\pm$0.05 & +0.7$\pm$0.8 &  U  \\          
34 & 18 20 33.060 & -16 11 21.59 & &  1.08$\pm$0.07 & 1.43$\pm$0.09 & $-$0.5$\pm$0.2 & U  \\
35 & 18 20 33.090 & -16 09 40.85 & &  0.95$\pm$0.19 & $\leq$0.20$\pm$0.05 & $\geq$2.9$\pm$0.6 & $0\rlap.{''}5$  \\
36 & 18 20 33.504 & -16 10 44.57 & &  0.49$\pm$0.08 &  0.40$\pm$0.08 & +0.4$\pm$0.5  & U  \\           
37 & 18 20 33.609 & -16 10 21.66 & &  0.22$\pm$0.05 & 0.29$\pm$0.06 & $-$0.5$\pm$0.6  & U  \\           
38 & 18 20 33.767 & -16 10 46.39 & &  0.26$\pm$0.06 & 0.27$\pm$0.08 & $-$0.1$\pm$0.7 & U \\           
\enddata
 \tablenotetext{a}{From the 8.46 GHz data, except for source 1, where the position
 is from the 4.96 GHz data. The positional error is estimated to be $\sim$$0\rlap.{''}1$. $^b$U = 
 Unresolved. $^c$Time variable sources. The flux densities listed are the average of the three
 observing sessions.}
 \end{deluxetable}

\clearpage

\begin{table}
\begin{center}
\small
\caption{Counterparts to compact radio continuum sources in M17.\label{tbl-4}}
\begin{tabular}{cccc}
\tableline\tableline
JVLA & & \multicolumn{2}{c}{Position$^a$} \\
\cline{3-4}  
Number & Counterpart$^b$ & $\alpha$(2000) & $\delta$(2000) \\
 \tableline
  3 & 2MASS J18201947-1613298 & 18 20 19.472 & -16 13 29.89  \\
  4 & Cl* NGC 6618 B 335 &  18 20 21.435  & -16 12 06.85 \\
  5 & Cl* NGC 6618 B 331 & 18 20 21.640 & -16 11 18.00 \\
  8 & Cl* NGC 6618 B 267 & 18 20 25.500 & -16 10 53.77 \\
 13 & Cl* NGC 6618 B 228 &  18 20 27.852 & -16 09 56.04 \\
 15 & [BFT2007] 471 & 18 20 28.260 & -16 13 17.91 \\
16 & Cl* NGC 6618 B 222 & 18 20 28.300 & -16 11 31.19 \\
 19 & CXOU J182029.4-161050 & 18 20 29.439 & -16 10 50.22 \\
  20 & CEN 1b =  [PCB2009] 2 & 18 20 29.811 & -16 10 45.67 \\
 21 & CEN 1a = [PCB2009] 1 & 18 20 29.900 & -16 10 44.40 \\
 22 & CXOU J182030.0-161034 & 18 20 30.030 & -16 10 34.81 \\
 23 & CXOU J182030.0-161039 & 18 20 30.080 & -16 10 39.97 \\
 25 & Cl* NGC 6618 B 179 & 18 20 30.346 & -16 11 43.69 \\
 27 & Cl* NGC 6618 B 174 & 18 20 30.448 & -16 10 53.09 \\
 28 & 2MASS J18203057-1611040 & 18 20 30.578 &  -16 11 04.10 \\
 29 & SSTGLMC G015.0618-00.6890 & 18 20 30.649 & -16 10 28.25 \\
 31 & Cl* NGC 6618 B 159 & 18 20 31.115 & -16 09 29.88 \\
 32 & 2MASS J18203152-1610326 & 18 20 31.523 &  -16 10 32.61 \\
 34 & Cl* NGC 6618 B 137 &  18 20 33.061 & -16 11 21.56 \\
 \tableline
 \end{tabular}
   \tablenotetext{a}{Position from the SIMBAD Astronomical Database. $^b$2MASS = 2 Micron All-Sky Survey;
   Cl* NGC 6618 = Star in cluster NGC 6618;
   B = Bumgardner (1992); BFT = Broos et al. (2007); CXOU = Chandra X-ray Observatory, Unregistered;
   PCB = Povich et al. (2009); SSTGLMC = Spitzer Space Telescope Galactic Legacy infrared Mid-Plane survey.}
\end{center}
\end{table}

\clearpage

\begin{table}
\begin{center}
\caption{Flux densities for CEN 1a and CEN 1b.\label{tbl-5}}
\begin{tabular}{ccccc}
\tableline\tableline
 & Frequency & 2011 Jun 16 & 2011 Jun 27 & 2011 Jul 28 \\
 Source & (GHz) & (mJy) & (mJy) & (mJy) \\
 \tableline
 CEN 1a & 4.96 & $\leq$0.27$\pm$0.09 & $\leq$0.21$\pm$0.07 & 2.69$\pm$0.09  \\
 CEN 1a & 8.46 & $\leq$0.26$\pm$0.05 & $\leq$0.24$\pm$0.06 & 2.58$\pm$0.05 \\
 CEN 1a & 22.46 & ---  & $\leq$0.42$\pm$0.10  & ---  \\
  & & & & \\
  CEN 1b & 4.96 & 1.17$\pm$0.10 & 1.17$\pm$0.11 & 1.15$\pm$0.09  \\
  CEN 1b & 8.46 & 0.81$\pm$0.05 & 0.66$\pm$0.06 & 0.80$\pm$0.05 \\
  CEN 1b & 22.46 & ---  & $\leq$0.42$\pm$0.10  & ---  \\
  \tableline
  \end{tabular}
  \end{center}
  \end{table}

\clearpage

\begin{table}
\begin{center}
\caption{Flux densities for NGC 6618 B 159.\label{tbl-6}}
\begin{tabular}{cccc}
\tableline\tableline
 Frequency & 2011 Jun 16 & 2011 Jun 27 & 2011 Jul 28 \\
  (GHz) & (mJy) & (mJy) & (mJy) \\
   \tableline
  4.96 & 1.89$\pm$0.09 & 1.67$\pm$0.13 & $\leq$0.21$\pm$0.06  \\
  8.46 & 3.82$\pm$0.07 & 1.61$\pm$0.06 & $\leq$0.18$\pm$0.04 \\
 22.46 & ---  & $\leq$1.56$\pm$0.42  & ---  \\
\tableline
\end{tabular}
\end{center}
\end{table}

\end{document}